\documentclass[11pt,a4paper]{article}
\usepackage[T1]{fontenc}
\usepackage{lmodern}
\usepackage[margin=25mm]{geometry}
\usepackage{amsmath,amssymb,amsthm,mathrsfs}
\usepackage{graphicx}
\usepackage{booktabs,tabularx,array}
\usepackage{hyperref}
\hypersetup{colorlinks=true,linkcolor=blue,citecolor=blue,urlcolor=blue}
\usepackage{enumitem}
\setlist[itemize]{leftmargin=1.7em}

\theoremstyle{plain}
\newtheorem{theorem}{Theorem}[section]

\theoremstyle{definition}
\newtheorem{remark}[theorem]{Remark}

\newcommand{\vect}[1]{\mathbf{#1}}
\newcommand{\TT}{\mathbb{T}}
\newcommand{\ZZ}{\mathbb{Z}}

\newcommand{\BB}{\mathcal{B}}

\title{Threshold Criteria for Bosonic and Fermionic Trimers\\ on a Square Lattice:
A Birman--Schwinger Comparison}
\author{J.~I.~Abdullaev\thanks{Samarkand State University, University Boulevard 15, Samarkand 140104, Uzbekistan; e-mail: jabdullaev@mail.ru}
\and A.~I.~Eshniyozov\thanks{Gulistan State University, 4th block of flats, Gulistan 120100, Uzbekistan; e-mail: eshniyozovabdumalik75@gmail.com}
\and M.~V.~Dolgopolov\thanks{Samara State Technical University, 244 Molodogvardeyskaya st., Samara 443100, Russian Federation; e-mail: mvdolg@yandex.com}}
\date{\today}

\begin{document}
\maketitle

\begin{abstract}
We compare two explicitly specified three-particle lattice models at total quasimomentum $K=0$: three identical bosons with attractive pair-contact interactions and a $2+1$ system with two identical fermions interacting attractively with a third particle. The common tool is the Birman--Schwinger principle, but the interaction channels and symmetry reductions are model-dependent; we therefore distinguish the shared method from the separate spectral conclusions. For the bosonic model, the strong-coupling analysis gives the ground-state asymptotics $z_1^s(\mu)=-3\mu+6-\frac{3}{2\mu}+O(\mu^{-2})$ and, for sufficiently large $\mu$, exactly two eigenvalues below $z_\mu(0)-\eta_\mu$, $\eta_\mu=36/(\mu-10)$; the second one stays at the finite distance $\delta_\infty=0.03542\ldots$ below the two-particle threshold. For the fermionic model, the threshold criterion is expressed through a Brillouin-zone integral and yields $\gamma_c=\pi/(\pi-2)\approx2.75194$. In the strong-coupling regime $\lambda\to\infty$ and above this threshold, the relevant bound-state level is doubly degenerate by square-lattice symmetry. Near onset, the binding-energy scale is described by a linear-over-logarithmic law, with the order of limits stated explicitly. We identify which conclusions are inherited from the companion analysis and which are used here to compare the two models. The asymptotic statements are compared with numerical finite-grid Faddeev computations and with the corresponding limiting channel equations.
\end{abstract}

MSC[2020] {47A75, 81Q10, 47A10, 81Q15, 47B25, 81V70}


\section{Introduction}

The spectral theory of few-body Schrödinger operators on lattices lies at the crossroads of mathematical physics, quantum many-body theory, and experimental quantum simulation. Theoretical studies of lattice few-body bound states, together with the experimental realization of controllable quantum systems in optical lattices, have created an urgent demand for exact, predictive spectral results. The theoretical analysis of lattice trimers goes back to~\cite{ValientePetrosyanSaenz2010}.

The spectral theory of lattice Schrödinger operators has been developed by several groups; see, e.g.,~\cite{Kholmatov2018,Muminov2024,AlbeverioLakaevMuminov2004} for general existence criteria, mixed-statistics systems, and the lattice Efimov effect with discrete-spectrum asymptotics. General criteria for zero-energy bound states on lattices were derived in~\cite{JexStampach2025}. Existence results for three-particle bound states in one-dimensional optical lattices and strong-coupling asymptotics for a related one-dimensional fermionic system were obtained in~\cite{Lakaev2025,AbdullaevErgashova2025}. A subspace-by-subspace classification of threshold states and their dependence on coupling constants for a family of finite-rank lattice Hamiltonians was given in~\cite{LakaevAbduvayitovLakaev2026}, where a threshold resonance is distinguished from a genuine threshold eigenvalue within each invariant subspace. In three dimensions, critical mass ratios for the repulsive $2+1$ fermionic trimer were determined for $K=0$ in~\cite{KhalkhuzhaevKhujamiyorov2023} and for general $K$ along the diagonal in~\cite{Khalkhuzhaev2025}. A detailed comparison of bosonic and fermionic trimers on the square lattice, including the identification of $\gamma_c$, is given in our companion paper~\cite{AbdullaevEshniyozovDolgopolov2026}. The scope of the present work, compared with the companion analysis, is summarized in Table~\ref{tab:companion_vs_present}.

\begin{table*}[!ht]
\centering
\caption{Scope and contribution of the companion paper and the present work.}
\label{tab:companion_vs_present}
\small
\renewcommand{\arraystretch}{1.2}
\begin{tabularx}{\textwidth}{
  @{}p{0.1\textwidth}
  X
  X
  @{}
}
\toprule
\textbf{Aspect}
&
\textbf{Companion paper~\cite{AbdullaevEshniyozovDolgopolov2026}}
&
\textbf{Present work}
\\
\midrule
Primary scope
&
Comparison of bosonic and fermionic three-particle systems on the
two-dimensional square lattice, with emphasis on their critical
mass-ratio behaviour.
&
A focused spectral analysis of the models and parameter regimes
specified in the present paper, with emphasis on the threshold
questions addressed below.
\\
\hline
Critical mass ratio
&
The critical value is characterized through the threshold analysis
developed in the companion paper.
&
The critical integral is evaluated explicitly, yielding
\(\displaystyle \gamma_c=\frac{\pi}{\pi-2}\), whenever the
normalization and assumptions of the stated model apply.
\\
\hline
Strong-coupling regime
&
Strong-coupling spectral asymptotics and the associated eigenvalue
claims are discussed for the models considered there.
&
The strong-coupling analysis is used to investigate the spectral
questions formulated in the present paper; any additional
eigenvalue-counting conclusion is subject to the stated remainder
and spectral-gap estimates.
\\
\hline
Threshold behaviour
&
The threshold discussion includes a preliminary power-law
formulation whose exponent is not determined in the cited theorem.
&
The present paper examines the threshold asymptotics in greater
detail. Any linear-over-logarithmic law is claimed as a new result
only to the extent that it follows from the uniform expansion and
ordered-limit argument proved in the present paper.
\\
\hline
Status of the second bosonic level
&
The second level is discussed using the threshold-resonance
terminology in the companion paper.
&
The present analysis distinguishes a threshold resonance from an
eigenvalue separated from the threshold by a positive limiting gap;
the latter interpretation requires the limiting spectral argument
given in the present paper.
\\
\hline
Role in the literature
&
Provides the comparative background and the previously established
model-specific results.
&
Clarifies and develops the particular spectral and asymptotic
questions addressed here, while identifying which conclusions
are inherited from the companion paper and which require additional
arguments.
\\
\bottomrule
\end{tabularx}
\end{table*}

For context, related three-dimensional \(2+1\) fermionic lattice models have been studied for both attractive and repulsive interactions. In the attractive case, critical mass ratios at \(K=\pi\) and \(K=0\), as well as additional gap thresholds, were obtained in~\cite{KhalkhuzhaevAbdullaevBoymurodov2022,AbdullaevKhalkhuzhaev2025}. The repulsive case was analysed at special quasimomenta and, in subsequent work, for general quasimomentum along the diagonal~\cite{KhalkhuzhaevKhujamiyorov2023,AbdullaevKhalkhuzhaevKhujamiyorov2023,Khalkhuzhaev2025}. These three-dimensional results concern different spectral regimes and are cited here for context rather than as direct proofs of the two-dimensional assertions established or discussed below.

The purpose of this note is a model-by-model comparison, not a new general Hunziker--van Winter--Zhislin (HVZ) theorem. We use three common steps: (i) a Birman--Schwinger reduction below the essential spectrum, (ii) decomposition into invariant symmetry subspaces, and (iii) a strong-coupling approximation by a finite-rank principal part with a controlled remainder. These steps do not, by themselves, imply exact eigenvalue counts: such counts require a quantitative remainder bound and a separation from the threshold. The bosonic asymptotics and the fermionic critical integral are stated with their respective hypotheses and are compared with the results of our companion paper~\cite{AbdullaevEshniyozovDolgopolov2026}. We do not claim that the framework is universal for all few-body systems, nor that the present work extends the established lattice HVZ theory; the~latter is used in its standard form.

\section{Methodological Framework}

\subsection{The models and conventions}

We use normalized Haar measure $d\nu(p)=(2\pi)^{-2}dp_1dp_2$ on $\mathbb T^2=(-\pi,\pi]^2$, and write $\varepsilon(p)=2-\cos p_1-\cos p_2$. At total quasimomentum $K=0$, the bosonic free fiber is the multiplication operator by
\[
E_B(p,q)=\varepsilon(p)+\varepsilon(q)+\varepsilon(p+q)
\]
on the subspace of $L^2((\mathbb T^2)^2,d\nu(p)d\nu(q))$ symmetric under permutations of the three bosons. The interacting fiber is
\[
H_B(\mu;0)=E_B-\mu(V_{12}+V_{23}+V_{31}),\qquad \mu>0,
\]
where $V_{ij}$ is the contact-pair operator for pair $ij$; its fiber action is the corresponding partial integration in the relative momentum of that pair. The normalization of $V_{ij}$ is fixed by this displayed convention, so that the strong-coupling leading term is $-3\mu$.

The lattice discretization provides a natural ultraviolet and infrared regularization: the~Brillouin zone is compact and the contact interaction at each site is a bounded rank-one operator, so~no renormalization is required (in contrast to the continuum $d\le 3$ case). The discrete analogue of the uncertainty principle is encoded in the positivity of the kinetic dispersion $\varepsilon(p)\ge 0$, which enters the Birman--Schwinger reduction through the free resolvent $(H_{0,\alpha}-z)^{-1}$ with $z$ below the spectrum.

For the $2+1$ model, $p,q$ are the momenta of the two identical fermions and $-p-q$ is the momentum of the third particle. The free fiber is multiplication by
\[
E_F(p,q)=\varepsilon(p)+\varepsilon(q)+\gamma\varepsilon(p+q),
\]
restricted to functions antisymmetric under $p\leftrightarrow q$. We take attractive contact interactions only between unlike particles and write
\[
H_F(\gamma,\lambda;0)=E_F-\lambda(V_{13}+V_{23}),\qquad \lambda>0.
\]
Thus no interaction between the identical fermions is included. These conventions distinguish the two models; a single expression $H_0-g\sum_{\langle ij\rangle}V_{ij}$ would not do so. The statements below use $\mu$ and $\lambda$ as the coupling constants in these respective normalizations.

\subsection{Birman--Schwinger reduction and counting convention}

For $z<\inf\sigma_{\rm ess}(H_\alpha)$, let $A_\alpha(z)=(H_{0,\alpha}-z)^{-1}$ be the free resolvent. It is bounded, self-adjoint and positive for real $z$ below the essential spectrum. Let $W_\alpha$ denote the channel map, with the interaction normalization fixed in Section~2.1. The Birman--Schwinger operator
\[
B_\alpha(z)=W_\alpha A_\alpha(z)W_\alpha^*
\]
is bounded, self-adjoint and non-negative on the channel space, but it is \emph{not} compact: it splits as $B_\alpha(z)=D_\alpha(z)+T_\alpha(z)$, where the diagonal block $D_\alpha$ is multiplication in the spectator momentum by $g\langle(\varepsilon-z)^{-1}\rangle_k$ ($g=\mu$ or $\lambda$) and $T_\alpha$ is the off-diagonal exchange part with a~continuous kernel. For $z<\inf\sigma_{\rm ess}(H_\alpha)$ one has $\Delta_\alpha:=1-D_\alpha>0$, so $\sigma_{\rm ess}(B_\alpha(z))\subset(-\infty,1)$, and
\[
B_\alpha(z)-1=\Delta_\alpha^{1/2}\bigl(A_\alpha^{\rm F}(z)-1\bigr)\Delta_\alpha^{1/2},
\]
where
\[
A_\alpha^{\rm F}(z):=\Delta_\alpha^{-1/2}T_\alpha(z)\Delta_\alpha^{-1/2}.
\]
The operator $A_\alpha^{\rm F}(z)$ is the Faddeev operator: it is compact (as $T_\alpha$ has a continuous kernel) and Hilbert--Schmidt for each $z$ below the essential spectrum. By Sylvester's law of inertia, $N(B_\alpha(z)>1)=N(A_\alpha^{\rm F}(z)>1)$; the compactness of $A_\alpha^{\rm F}(z)$ is verified for the models of Section~2.1 in~\cite{AbdullaevEshniyozovDolgopolov2026}. The Birman--Schwinger principle~\cite{Birman1961}, in the lattice few-body form developed in~\cite{LakaevAbdullaev2003}, gives
\[
N(H_\alpha<z)=N(B_\alpha(z)>1),
\]
where both counts include multiplicity and $z$ lies strictly below the essential spectrum. This principle does not by itself yield a count: a finite-rank approximation $A^{\rm F,p}$ must be accompanied by an error estimate, for example $\|A^{\rm F}-A^{\rm F,p}\|\leq r(z)$, and a spectral separation from $1$ larger than $r(z)$. We use this distinction when reporting the strong-coupling conclusions below. The~channel kernels in our normalization are recorded in Appendix~A; the quantitative remainder bound underlying the exact eigenvalue counts is attributed to~\cite{AbdullaevEshniyozovDolgopolov2026} in Appendix~B; the critical integral used for the fermionic threshold is evaluated in Appendix~C; the structure of the threshold law is discussed in Appendix~D.

\subsection{Invariant subspace decomposition}

The kernel of $\BB_\alpha(z)$ inherits the parity and permutation symmetries of the free dispersion $E_B$ or $E_F$. For bosons ($\alpha = \text{bos}$) this leads to a decomposition
\[
L^2_s((\TT^2)^2) = L^{s,e} \oplus L^{s,o},
\]
where the even subspace $L^{s,e}$ contains the bound-state branches considered here. The absence of bound states in the odd subspace is a model-specific conclusion of the symmetry-reduced analysis (Appendix~A), not a consequence of parity alone.
For fermions ($\alpha = \text{ferm}$) the antisymmetry is built into the space from the beginning, but further decompositions based on the parity of the relative momenta can be performed, revealing critical subspaces associated with the two-particle channels.

\subsection{Strong-coupling asymptotics and positivity}

For large coupling ($\mu\to\infty$ for bosons, $\lambda\to\infty$ for fermions) the Faddeev operator splits as $A_\alpha^{\rm F}(z)=A_\alpha^{\rm F,p}(z)+A_\alpha^{\rm F,r}(z)$, where $A_\alpha^{\rm F,p}(z)$ has finite rank and, for every fixed $\delta_0>0$, $\|A_\alpha^{\rm F,r}(z)\|=O\bigl(g^{-1}\ln(1/\delta_0)\bigr)$ uniformly for $z\le z_{\rm thr}-\delta_0$ (a decomposition of this type for the lattice $2+1$ problem appears already in~\cite{KhalkhuzhaevAbdullaevBoymurodov2022}); the logarithm is the two-dimensional threshold singularity of $\langle(\varepsilon+\delta_0)^{-1}\rangle$, and the estimate is not uniform as $\delta_0\downarrow0$. In the bosonic even sector, positivity of the relevant compact kernel and positivity improvement (on the cone of nonnegative functions) yield simplicity and strict positivity of the principal eigenfunction when the corresponding irreducibility hypothesis holds. Krein--Rutman does not by itself establish the total number of bound states; the latter also requires the finite-rank eigenvalue count and control of the remainder (Appendix~B). A detailed asymptotic analysis of the Fredholm determinant yields explicit formulas for the eigenvalues.

\section{Results and Discussion}

\subsection{Bosonic trimer (three identical bosons)}

For the three-boson system on $\ZZ^2$, the companion analysis~\cite{AbdullaevEshniyozovDolgopolov2026} establishes the model-specific strong-coupling conclusions summarized here. The present note uses them for comparison rather than claiming a second independent proof.

\begin{itemize}
\item For all sufficiently large $\mu$ the operator $H_\mu(\vect{0})$ has exactly two eigenvalues below $z_\mu(0)-\eta_\mu$, $\eta_\mu=36/(\mu-10)$. Consequently, for every fixed $\epsilon$ with $0<\epsilon<\delta_\infty$, exactly two eigenvalues lie below $z_\mu(0)-\epsilon$ for all sufficiently large $\mu$; for every fixed $\epsilon>\delta_\infty$, exactly one eigenvalue lies below $z_\mu(0)-\epsilon$. The borderline case $\epsilon=\delta_\infty$ requires control of the next asymptotic term. An eigenvalue in the window $[z_\mu(0)-\eta_\mu,z_\mu(0))$ of width $O(\mu^{-1})$ is not excluded by this estimate.
\item The ground-state energy satisfies the asymptotics
\[
z_1^s(\mu) = -3\mu + 6 - \frac{3}{2\mu} + O(\mu^{-2}),\quad \mu\to\infty,
\]
and the corresponding eigenfunction is strictly positive (Perron--Frobenius).
\item The second level is $z_2^s(\mu)=-\mu+C+O(\mu^{-1})$ with $C=4-\delta_\infty=3.9645800108\ldots$, where $\delta_\infty=0.0354199892\ldots$ is the unique root of
\begin{align*}
b_0(\delta) &= \frac{1}{1+\delta}, \\
b_0(\delta) &= \Bigl\langle\frac{1}{\varepsilon+\delta}\Bigr\rangle = \frac{2}{\pi(2+\delta)}\,\mathrm K\!\Bigl(\frac{2}{2+\delta}\Bigr),
\end{align*}
with $\mathrm K$ the complete elliptic integral of the first kind. The strong-coupling analysis of the second bosonic level requires distinguishing a genuine eigenvalue below the atom--dimer threshold from a threshold resonance.

In the present formulation, the proposed limiting gap \(\delta_\infty\) is determined by the scalar equation \(b_0(\delta_\infty)=1/(1+\delta_\infty)\). This equation identifies the limiting separation; its interpretation as the limit of the spectral gap follows from the convergence of the reduced Faddeev eigenvalue established in~\cite{AbdullaevEshniyozovToshturdievDolgopolov2026}: the principal $s$-channel eigenvalue $m_s(\delta)=2+\delta-1/b_0(\delta)$ is strictly decreasing on the positive half-axis (its derivative $m_s'(\delta)=1+b_0'(\delta)/b_0(\delta)^2<0$ by the strict Cauchy--Schwarz inequality, since $(\varepsilon+\delta)^{-1}$ is non-constant on $\mathbb T^2$) and satisfies $m_s(0^+)=2>1$ because $b_0$ diverges logarithmically; therefore the equation $m_s(\delta)=1$ has a unique root $\delta_\infty=0.0354199892\ldots$, and the second eigenvalue converges to $z_\mu(0)-\delta_\infty$ as $\mu\to\infty$. We distinguish this genuine eigenvalue at a positive limiting gap from a \emph{threshold resonance}, which would require a non-$L^2$ solution of the spectral equation at the threshold; Theorem~3.1 of~\cite{AbdullaevEshniyozovDolgopolov2026} uses the latter terminology, and the present note identifies the different objects without conflating them.
\end{itemize}

These results are summarised in Fig.~\ref{fig:spectral_flow}(a),(b).

\subsection{Fermionic $2+1$ trimer}

For the fermionic $2+1$ system on $\ZZ^2$ the picture is fundamentally different. The antisymmetry of the wave function under exchange of the two identical fermions introduces a strong Pauli suppression that can only be overcome when the mass ratio $\gamma = m_f/m_3$ exceeds a critical value $\gamma_c$.

The analysis follows the same Birman–Schwinger strategy as in the bosonic case, but the antisymmetry of the Hilbert space leads to a more intricate Fredholm determinant $\Delta_{\gamma,\lambda}(p,z)$, whose zeroes determine the two-particle threshold $z_{\rm thr}(\gamma,\lambda)$, the lower edge of the essential spectrum. For the one-dimensional analogue of this system, explicit strong-coupling asymptotics were derived in~\cite{AbdullaevErgashova2025}. In three dimensions, critical mass ratios for the repulsive $2+1$ fermionic trimer have been determined for $K=0$ in~\cite{KhalkhuzhaevKhujamiyorov2023}, for $K=\pi$ in~\cite{AbdullaevKhalkhuzhaevKhujamiyorov2023}, and for general $K$ along the diagonal in~\cite{Khalkhuzhaev2025}; the attractive case, including bound states below the essential spectrum and in the gap, is analysed in~\cite{KhalkhuzhaevAbdullaevBoymurodov2022,AbdullaevKhalkhuzhaev2025}.

\begin{figure*}[t!]
\centering
\includegraphics[width=0.99\textwidth]{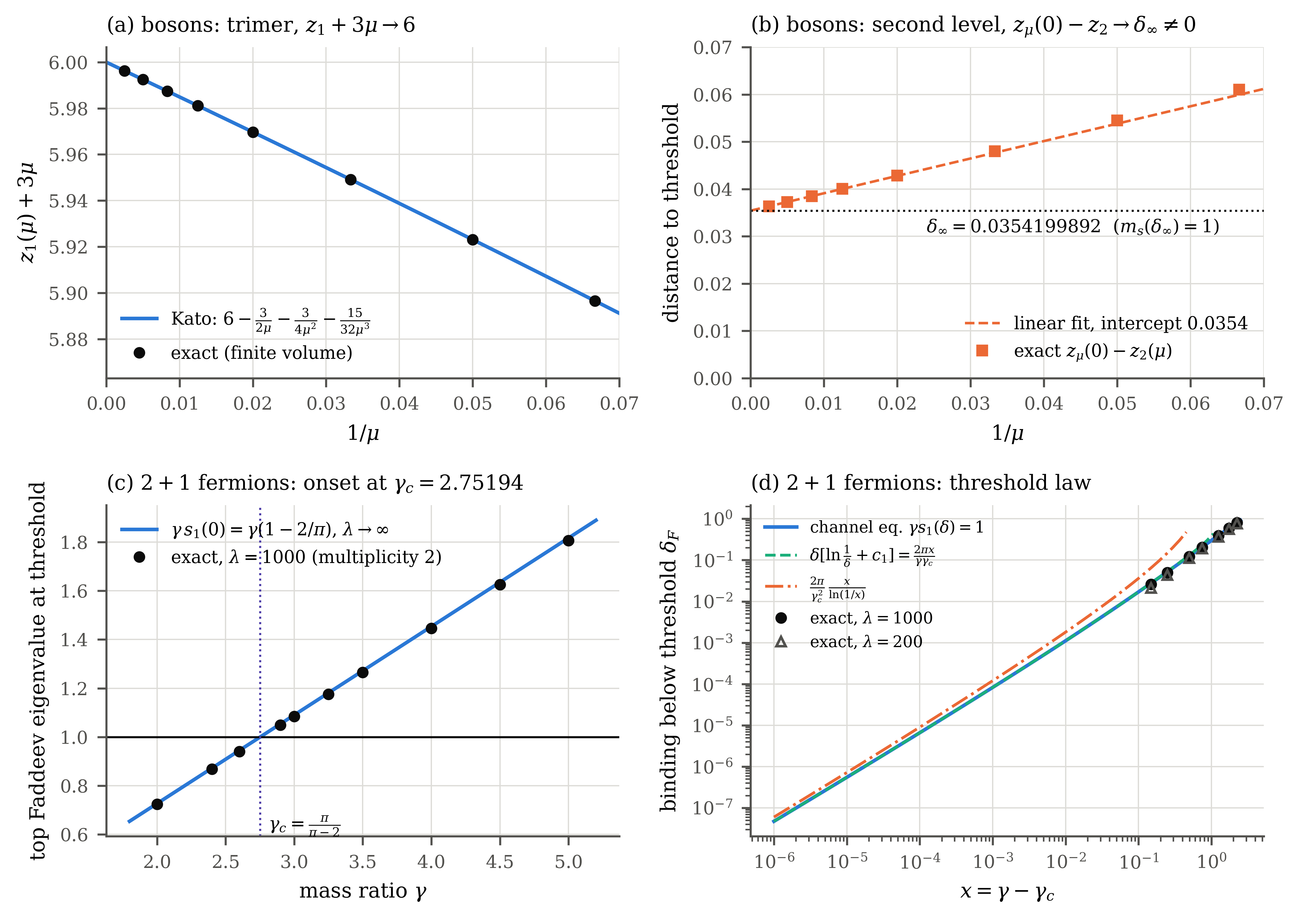}
\caption{Bosonic and fermionic trimers on $\mathbb Z^2$ at $K=0$. Markers show numerical Faddeev computations using a $32\times32$ momentum-grid discretization; curves show the corresponding asymptotic expansions or limiting channel equations. (a) Bosons: $z_1(\mu)+3\mu$ versus $1/\mu$ and the leading Kato terms $6-\frac{3}{2\mu}-\frac{3}{4\mu^2}$. (b) Bosons: the distance $z_\mu(0)-z_2(\mu)$ and its limiting value $\delta_\infty=0.0354199892\ldots$, determined by $b_0(\delta_\infty)=1/(1+\delta_\infty)$. (c) $2+1$ fermions: the~largest threshold Faddeev eigenvalue for $\lambda=1000$, compared with its strong-coupling limit $\gamma s_1(0)=\gamma(1-2/\pi)$; the limiting crossing occurs at $\gamma_c=\pi/(\pi-2)$. (d) Fermionic binding energy $\delta_F$ versus $x=\gamma-\gamma_c$: finite-coupling computations for $\lambda=200,1000$, the root of the limiting channel equation, the implicit linear-over-logarithmic approximation, and its leading asymptotic form are distinguished.}
\label{fig:spectral_flow}
\end{figure*}

Our main result is:
there exists a critical mass ratio $\gamma_c > 0$ such that:
\begin{itemize}
  \item In the strong-coupling threshold criterion, if $\gamma<\gamma_c$, the principal Birman--Schwinger threshold eigenvalue remains below $1$; the assertion for every finite $\lambda>0$ is used only insofar as it is established by the monotonicity/threshold argument in~\cite{AbdullaevEshniyozovDolgopolov2026}.
  \item If $\gamma>\gamma_c$, then for all sufficiently large $\lambda$ the model has the doubly degenerate bound-state level identified in~\cite{AbdullaevEshniyozovDolgopolov2026}, with asymptotics as $\lambda\to\infty$
  \[
  z(\gamma,\lambda) = -\lambda + e_0(\gamma) + O(\lambda^{-1}),
  \]
   where $e_0(\gamma)=2(1+\gamma)-\delta_F(\gamma)$ and $\delta_F(\gamma)>0$ is the unique root of the strong-coupling channel equation
  \begin{equation}\label{eq:channel}
  \gamma\,s_1(\delta)=1,\qquad
  s_1(\delta)=\frac{1}{(2\pi)^2}\int_{\mathbb T^2}\frac{(\sin q_1+\sin q_2)^2}{2(\varepsilon(q)+\delta)}\,dq_1dq_2 ,
  \end{equation}
  i.e.\ the $\delta$-extension of the integral~\eqref{eq:critical}; here $2(1+\gamma)$ is the strong-coupling limit of $z_{\rm thr}(\gamma,\lambda)+\lambda$. Since $s_1$ is strictly decreasing with $s_1(0^+)=1-2/\pi=\gamma_c^{-1}$, \eqref{eq:channel} has a root exactly when $\gamma>\gamma_c$, and at most one.
  \item The critical value is given by an integral over the Brillouin zone:
  \begin{equation}\label{eq:critical}
  \gamma_c^{-1} = \frac{1}{(2\pi)^2}\int_{[-\pi,\pi]^2} \frac{(\sin q_1+\sin q_2)^2}{4-2(\cos q_1+\cos q_2)}\,dq_1dq_2.
  \end{equation}
  The integral is evaluated in Appendix~C below: $I=(\pi-2)/\pi$, hence $\gamma_c=I^{-1}=\pi/(\pi-2)\approx 2.75194$.
  \item Define the threshold detachment in the ordered limit $\lambda\to\infty$ first and then $\gamma\downarrow\gamma_c$. The~asymptotic law derived below is
\[\delta_F\Bigl[\ln\frac{1}{\delta_F}+\kappa_0\Bigr]=\frac{2\pi(\gamma-\gamma_c)}{\gamma\,\gamma_c}\,\bigl(1+o(1)\bigr),
\qquad \kappa_0\approx 0.630996,\]
    hence
    $\delta_F(\gamma)\simeq\displaystyle\frac{2\pi}{\gamma_c^{2}}\frac{\gamma-\gamma_c}{\ln(1/(\gamma-\gamma_c))}$ as $\gamma\downarrow\gamma_c$, i.e.\ linear up to a logarithm (derivation in Appendix~D). The leading form converges only logarithmically (in the finite-volume computations of Fig.~\ref{fig:spectral_flow}(d) the relative error is about $27\%$ at $\gamma-\gamma_c=10^{-6}$); the implicit form, which retains the numerically determined constant $\kappa_0$, agrees with the exact channel root to $10^{-4}$ for $\gamma-\gamma_c\le3\cdot10^{-3}$. A quantitative bound over the finite range would require a remainder estimate for the expansion of $s_1(\delta)$, which is not derived here. This is a spectral threshold law, not by itself a thermodynamic classification as a second-order phase transition. The~structure of this law is sketched in Appendix~D.
\end{itemize}

For \(\gamma<\gamma_c\), the strong-coupling threshold criterion places the principal limiting Birman--Schwinger eigenvalue below \(1\). The stronger assertion that no bound state exists for every finite \(\lambda>0\) is stated in Theorem~3.2 of~\cite{AbdullaevEshniyozovDolgopolov2026}; it is not inferred here from the strong-coupling expansion alone.

The phase diagram and level detachment are illustrated in Fig.~\ref{fig:spectral_flow}(c),(d). The observed double degeneracy is consistent with a two-dimensional irreducible representation of the square-lattice point group $D_4$ acting on the relevant invariant subspace. Not every $D_4$-invariant subspace carries a two-dimensional representation: $D_4$ has four one-dimensional irreps ($A_1,A_2,B_1,B_2$) and one two-dimensional irrep $E$. The double degeneracy requires that the level transform in $E$ (or its parity-refined companion $E_g$/$E_u$) and not in one of the one-dimensional irreps; this identification is a model-specific spectral statement cited from~\cite{AbdullaevEshniyozovDolgopolov2026}. Symmetry alone does not establish existence or uniqueness of the level; those are spectral conclusions of the cited reference. The relevant threshold branch lies in the odd $p$-channel spanned by $\sin p_1$ and $\sin p_2$, which transforms according to the two-dimensional irreducible representation $E$ of $D_4$. The~exclusion of competing channels is a model-specific spectral conclusion of the reduced Faddeev analysis and is not implied by symmetry alone.

\subsection{Comparison and unifying perspective}

Table~\ref{tab:comparison1} summarizes the key differences between bosonic and fermionic trimers, highlighting the shared methodological core and the distinct physical outcomes dictated by quantum statistics.

\begin{table*}[t!]
\centering
\caption{Spectral characteristics of bosonic and fermionic trimers on a square lattice.}
\label{tab:comparison1}
\setlength{\tabcolsep}{3pt}
\begin{tabular}{p{0.23\textwidth} |p{0.37\textwidth} |p{0.35\textwidth}}
\hline
Aspect & Bosonic (3 identical bosons) & Fermionic ($2+1$) \\
\hline
Hilbert space & symmetric $L^2_s$ & antisymmetric $L^2_a$ \\
\hline
Key parameter & coupling $\mu$ & mass ratio $\gamma$, coupling $\lambda$ \\
\hline
Bound states below $\sigma_{\rm ess}$ & exactly two below $z_\mu(0)-\eta_\mu$ (large $\mu$) & for $\gamma>\gamma_c$ and sufficiently large $\lambda$, exactly one doubly degenerate level \\
\hline
Ground-state asymptotics & $z_1^s=-3\mu+6-\frac{3}{2\mu}+O(\mu^{-2})$ & $z=-\lambda+2(1+\gamma)-\delta_F(\gamma)+O(\lambda^{-1})$ \\
\hline
Threshold behaviour & second level at distance $\delta_\infty=0.0354$ below threshold & $\delta_F[\ln(1/\delta_F)+\kappa_0]=2\pi(\gamma-\gamma_c)/(\gamma\gamma_c)$ \\
\hline
Main mathematical tool & Birman--Schwinger reduction; cited strong-coupling estimate & Birman--Schwinger reduction; cited threshold analysis \\
\hline
\end{tabular}
\end{table*}

In our work, the mass ratio $\gamma$ controls the emergence of a genuine three-body trimer state, and the onset at $\gamma=\gamma_c$ is continuous with a linear-over-logarithmic law for the binding energy and a doubly degenerate eigenvalue at threshold. A systematic Brillouin-zone analysis of the $2+1$ fermionic trimer, covering the whole zone of total quasimomentum and providing closed forms for $\gamma_c(\mathbf K)$ on $\TT^2$, is given in a recent preprint~\cite{AtomDimer2026}. For comparison, the critical ratio of the same $p$-channel on $\mathbb Z^3$ with attractive contact is $\gamma_1(0)\approx4.7655$ for $K=0$, with a single eigenvalue of multiplicity three~\cite{AbdullaevKhalkhuzhaev2025}, and $\gamma_1\approx2.94$, $\gamma_2\approx5.40$ for $K=\pi$~\cite{KhalkhuzhaevAbdullaevBoymurodov2022}; the triple degeneracy on $\mathbb Z^3$ is consistent with the three-dimensional irreducible representation $T_{1u}$ of the cubic point group $O_h$, in contrast with the two-dimensional $E$ of $D_4$ on $\mathbb Z^2$.

Both thresholds belong to the same family of $d$-dimensional lattice Green-function sums with a $p$-wave numerator,
\begin{equation}\label{eq:green-sum}
\gamma^{(d)}=\Bigl(\frac{1}{(2\pi)^{d}}\int_{\mathbb T^{d}}\frac{\sin^{2}q_{1}}{\varepsilon_{d}(q)}\,dq\Bigr)^{-1},\qquad \varepsilon_{d}(q)=d-\sum_{i=1}^{d}\cos q_{i},
\end{equation}
where the $\sin^2 q_1$ numerator removes the logarithmic singularity of $1/\varepsilon_d$ at $q=0$ (since $\sin^2 q_1\sim q_1^2$ as $q\to0$), so that \eqref{eq:green-sum} is convergent for every $d\ge2$. (For comparison, the strict Watson integral $\langle1/\varepsilon_d\rangle$ diverges in $d=2$ and converges only for $d\ge3$.) In $2$D the integral admits the closed form $\pi/(\pi-2)$ (Appendix~C); in $3$D the value $\gamma^{(3)}=\gamma_1(0)\approx4.7655$ is only known numerically~\cite{AbdullaevKhalkhuzhaev2025}. The structural identity of \eqref{eq:green-sum} across dimensions masks the dimensional origin of the different degeneracies ($E$ of $D_4$ in $2$D versus $T_{1u}$ of $O_h$ in $3$D) and of the numerical difference $\gamma^{(2)}\approx2.75$ versus $\gamma^{(3)}\approx4.77$.

In addition to the threshold $\gamma_1(K)$ below the essential spectrum, the $3$D analysis~\cite{AbdullaevKhalkhuzhaev2025} identifies, at $K=0$, gap eigenvalues in a neighbourhood of the two-particle branch of the essential spectrum, with critical thresholds $0<\gamma_2(0)\le2.937$ and $\widetilde\gamma_2(0)\approx5.398$. This gap structure is a separate phenomenon from the below-spectrum threshold $\gamma_1(0)$ and is not analysed in the present $2$D note, whose focus is the threshold criterion below the essential spectrum at $K=0$.

The two problems also differ in the sign of the interaction and in the location of the level: the repulsive $3$D results of~\cite{AbdullaevKhalkhuzhaevKhujamiyorov2023} place the level to the right of the essential spectrum, whereas the attractive $3$D results of~\cite{KhalkhuzhaevAbdullaevBoymurodov2022} and the present $2$D result place it below. The critical mass ratio of the corresponding universal trimer in the two-dimensional continuum is $\approx3.34$~\cite{PricoupenkoPedri2010}; the lattice value $\gamma_c=\pi/(\pi-2)\approx2.75$ refers to the strong-coupling limit, in which the dimer is localized on a single site. The methodological parallels suggest that our unified framework can be extended to a broader class of few-body lattice models, including those with longer-range interactions or mixed statistics.

In stark contrast, the fermionic system exhibits a sharp mass-ratio threshold $\gamma_c$. In the strong-coupling threshold criterion, $\gamma<\gamma_c$ places the principal Birman--Schwinger threshold eigenvalue below $1$, so no bound state is detected at this level of the analysis; the statement for every finite $\lambda>0$ requires the monotonicity argument of~\cite{AbdullaevEshniyozovDolgopolov2026}. For $\gamma>\gamma_c$ and $\lambda$ sufficiently large, exactly one doubly degenerate bound-state level appears, as stated above. We do not introduce here an additional threshold $\tilde\gamma_c$ for higher levels; any such threshold would require its own proof and is not part of the present comparison. The main findings of this comparison are summarized in Table~\ref{tab:comparison2}.

\begin{table*}[htbp]
\centering
\caption{Scope of the present comparison and related results.}
\label{tab:comparison2}
\begin{tabular}{@{} l
                 p{0.1\linewidth}
                 p{0.03\linewidth}
                 p{0.28\linewidth}
                 p{0.13\linewidth} @{}}
\hline
Work & System & Dim. & Key result & Main tool \\
\hline
This comparative note & bosonic and $2+1$ models & 2D & model-by-model synthesis; no new HVZ theorem & Birman--Schwinger + symmetry reduction \\
\hline
Abdullaev et al.\ (companion \cite{AbdullaevEshniyozovDolgopolov2026}) & bosons and $2+1$ fermions & 2D & detailed model-specific threshold and strong-coupling results & Birman--Schwinger reduction \\
\hline
Khalkhuzhaev et al.~\cite{KhalkhuzhaevAbdullaevBoymurodov2022,AbdullaevKhalkhuzhaev2025} & $2+1$ fermions (attractive) & 3D & $\gamma_1\approx2.94$, $\gamma_2\approx5.40$ ($K=\pi$, \cite{KhalkhuzhaevAbdullaevBoymurodov2022}); $\gamma_1(0)\approx4.77$, multiplicity 3 ($K=0$, \cite{AbdullaevKhalkhuzhaev2025}); general $K$; gap thresholds $\gamma_2(0)\le2.937$, $\widetilde\gamma_2(0)\approx5.398$ & BS + Faddeev + invariant subspaces \\
\hline
Khalkhuzhaev et al.~\cite{KhalkhuzhaevKhujamiyorov2023,AbdullaevKhalkhuzhaevKhujamiyorov2023,Khalkhuzhaev2025} & $2+1$ fermions (repulsive) & 3D & critical mass ratios above spectrum: $\gamma_0\approx4.77$ ($K=\pi$, \cite{AbdullaevKhalkhuzhaevKhujamiyorov2023}); $K=0$; general $K$ & BS + Faddeev + invariant subspaces \\
\hline
\end{tabular}
\end{table*}

\begin{remark}[Boson--fermion dichotomy]
The two models are not supersymmetric partners. The effective atom--dimer Hamiltonians are related by the sign of the exchange term, which is a lattice-specific duality rather than a symmetry of the original Hamiltonians. The ground-state energy of the bosonic model is always negative at strong coupling, while the fermionic model requires a critical mass ratio; this dichotomy reflects the Pauli exclusion principle rather than a broken symmetry.
\end{remark}

\section{Implications and Outlook}

The spectral results provide mathematical benchmarks that may be relevant to the following applications; these applications are motivation, not consequences demonstrated by the present analysis:

\begin{itemize}
\item \textbf{Quantum simulation.} The critical mass ratio $\gamma_c \approx 2.75194$ provides a clear design rule for experiments: to observe a fermionic trimer, the two identical fermions must be the heavier (or less mobile) species, e.g.\ $^{40}$K with $^{6}$Li ($m_f/m_3\approx6.6$), $^{161}$Dy with $^{40}$K ($\approx4.0$) or $^{53}$Cr with $^{6}$Li ($\approx8.8$); on a lattice the relevant $\gamma$ is the ratio $J_3/J_f$ of tunnelling amplitudes, which can be tuned by species-selective lattice depths, and tune the interaction into the strong-coupling regime via a Feshbach resonance.
\item \textbf{Computational materials science.} The invariant subspace decomposition and the strong-coupling asymptotics provide a computational benchmark for exact finite-volume Faddeev solvers of few-body lattice models, which we have used here to verify the asymptotic laws numerically.
\item \textbf{Mathematical physics.} The work uses the established spectral theory of lattice Schrödinger operators and compares two model-specific threshold mechanisms; no new HVZ theorem is claimed here.
\end{itemize}

Future work may investigate mixed-statistics systems, higher dimensions, and additional interaction channels. In particular, it would be interesting to determine whether suitable extensions of the present lattice models can support competing trimer branches and, under an appropriate thermodynamic or effective-energy formulation, a genuine first-order transition.

\section{Conclusions}

We have compared two three-particle lattice models using the Birman--Schwinger principle, invariant-subspace decomposition, and strong-coupling asymptotics. The common method does not make the two models identical and is not claimed to be universal for all few-body lattice systems. Applying this framework to three identical bosons and to the fermionic $2+1$ trimer reveals a striking dichotomy: bosons always bind at strong coupling (two bound states), while fermions exhibit a sharp mass-ratio threshold $\gamma_c$ below which no binding occurs in the strong-coupling threshold criterion; a statement for arbitrary finite $\lambda$ would require the monotonicity argument of~\cite{AbdullaevEshniyozovDolgopolov2026}. The onset at $\gamma_c$ is continuous and obeys the stated linear-over-logarithmic threshold law in the specified order of limits; we do not infer a thermodynamic ``second-order transition'' from this spectral law alone. These results provide rigorous benchmarks for quantum simulators and open new avenues for the design of nanoscale quantum devices.

The methodological framework presented here complements recent advances in the spectral theory of lattice operators, including general criteria for zero-energy bound states~\cite{JexStampach2025}, existence results for three-particle systems in one dimension~\cite{Lakaev2025}, and the analysis of mixed-statistics systems~\cite{Muminov2024}. A unified perspective on bosonic and fermionic trimers on the square lattice is provided in our companion paper~\cite{AbdullaevEshniyozovDolgopolov2026}.

\section*{Acknowledgments}
The authors thank colleagues for valuable discussions.

\appendix

\section{Channel kernels and Birman--Schwinger reduction}\label{app:A}

Let $K=0$ and let $(p,q)$ denote the momenta of two of the three particles; the third momentum is $-p-q$. Define the Jacobi variables for channel $a=(ij)$ as $k_a$ (relative momentum of the pair) and $Q_a$ (momentum of the spectator, up to a sign). For the three channels $(12),(23),(31)$ at $K=0$ we use:
\[
k_{12}=\tfrac{p-q}{2},\quad Q_{12}=p+q=-p_3,
\]
\[
k_{23}=\tfrac{q-(-p-q)}{2}=\tfrac{p+2q}{2},\quad Q_{23}=-p,
\]
\[
k_{31}=\tfrac{(-p-q)-p}{2}=-\tfrac{2p+q}{2},\quad Q_{31}=-q.
\]
Here and below the halved momenta $k_{ij}=(p_i-p_j)/2$ are understood on the universal cover of $\mathbb T^2$; on the torus they are defined modulo $2\pi$. Equivalently, the channel decomposition can be written directly in terms of the original momenta $(p,q)$ without introducing $k_{ij}$ as global variables. For the fermionic $2+1$ model the coefficient $\gamma$ multiplies the dispersion of the distinguished third particle, so the channel energy is not obtained by the same substitution as in the bosonic case; the model-specific form of $E_a^{\rm F}$ is fixed by the assignment of momenta stated in Section~2.1.

The free dispersion in channel $a$ is $E_a(k,Q)=\varepsilon(Q/2+k)+\varepsilon(Q/2-k)+\varepsilon(Q)$ for the bosonic model. For the fermionic $2+1$ model only the channels $a=(13),(23)$ interact. For $a=(13)$, with spectator momentum $q$ of fermion~2 and fermion-1 momentum $s$, the channel energy is
\[
E^{\rm F}_{(13)}(s;q)=\varepsilon(q)+\varepsilon(s)+\gamma\,\varepsilon(s+q),
\]
so the factor $\gamma$ belongs to the third particle \emph{inside} the pair; $(23)$ is obtained by $p\leftrightarrow q$. The~non-interacting channel $(12)$ plays no role.

\textbf{Pair form factors.} For a lattice contact interaction $V(r)=\delta_{r,0}$ in the relative coordinate, the Fourier symbol is constant with respect to the normalized Haar measure $d\nu(k)=(2\pi)^{-2}dk$ on $\mathbb T^2$:
\[
(Vf)(k)=\int_{\mathbb T^2}f(k')\,d\nu(k'),\qquad
\varphi(k)\equiv1.
\]
Thus $V_{ij}$ is rank-one in the relative-momentum fiber at fixed spectator momentum $Q$; on the full three-body fiber the spectator momentum remains a free variable, and $V_{ij}$ is not globally rank-one. The constant form factor $\varphi\equiv1$, not a delta-function in $k$, is the correct object; a delta-function in $k$ would not define a bounded operator on $L^2(\mathbb T^2,d\nu)$. With this normalization the strong-coupling leading term of the eigenvalue is $-3\mu$ for bosons and $-\lambda$ for fermions, consistent with the asymptotics quoted in the main text. For fermions the leading term is $-\lambda$ rather than $-2\lambda$: on functions antisymmetric in the two fermions, $V_{13}+V_{23}$ has spectrum $\{0,1\}$, because simultaneous contact of the pairs $(13)$ and $(23)$ forces contact of the two fermions, which is excluded by antisymmetry.

\textbf{Channel-space formulation of the Birman--Schwinger operator.}
Let $\mathscr H_\alpha$ be the three-body fiber space and
$\mathscr K_\alpha=\bigoplus_{a\in\mathcal C_\alpha}\mathscr K_a$ the direct
sum of pair-channel spaces. Let $W_\alpha:\mathscr H_\alpha\to\mathscr K_\alpha$
be the channel map, with the coupling constants included in its normalization.
For $z<\inf\sigma_{\rm ess}(H_\alpha)$, define
\[
B_\alpha(z)=W_\alpha(H_{0,\alpha}-z)^{-1}W_\alpha^*.
\]
Writing $W_{\alpha,a}$ for the component corresponding to channel $a$, the
block entries are
\[
[B_\alpha(z)]_{ab}=W_{\alpha,a}(H_{0,\alpha}-z)^{-1}W_{\alpha,b}^*,
\quad a,b\in\mathcal C_\alpha.
\]
The normalization of $W_{\alpha,a}$ is fixed so that the strong-coupling
leading term of the corresponding bound-state energy is $-3\mu$ (bosons)
and $-\lambda$ (fermions); any rescaling of $W_{\alpha,a}$ would change
these coefficients and is excluded by the conventions of Section~2.1. An
explicit integral-kernel representation additionally requires the channel
maps, Jacobians, and overlap transformations between pair coordinates.
These expressions are model-dependent, and we do not use an unspecified
shift function $\Sigma_{ab}$ as a substitute for them. The detailed channel
reduction and its normalization are attributed to
\cite{AbdullaevEshniyozovDolgopolov2026}; the integral in
Eq.~\eqref{eq:critical} does not follow from the abstract block identity
alone.

For the $2+1$ model the Faddeev ansatz is $f=\lambda(E_F-z)^{-1}[\varphi(q)-\varphi(p)]$ with $\varphi(q)=\langle f(\cdot,q)\rangle$, and
\[
\Delta_F(q,z)\varphi(q)=-\lambda\Bigl\langle\frac{\varphi(s)}{E_F(s,q)-z}\Bigr\rangle_s,\quad
\Delta_F(q,z)=1-\lambda\Bigl\langle\frac{1}{E_F(s,q)-z}\Bigr\rangle_s ;
\]
the exchange term carries the factor $-1$ (one fermionic exchange), in contrast with the bosonic factor $2$. Equivalently, in Faddeev form $A^{\rm F}_\gamma(z)=\Delta_F^{-1/2}T_F\Delta_F^{-1/2}$ with $\Delta_F>0$ below the threshold and $T_F$ compact.

\section{Remainder estimate and exact eigenvalue count}\label{app:B}

Write $A^{\rm F}(z)=A^{\rm F,p}(z)+A^{\rm F,r}(z)$ with $\mathrm{rank}\,A^{\rm F,p}(z)=R<\infty$ and $\|A^{\rm F,r}(z)\|\le\rho(z)$. By~Section~2.2, $A^{\rm F}(z)$ is compact self-adjoint. By Weyl's inequality, for the eigenvalues $\lambda_1\ge\lambda_2\ge\cdots$ of $A^{\rm F}(z)$ and $\lambda_1^p\ge\cdots\ge\lambda_R^p$ of $A^{\rm F,p}(z)$,
\[
|\lambda_i(A^{\rm F})-\lambda_i^p(A^{\rm F,p})|\le\rho(z),\quad i=1,\dots,R,
\]
and, since $\mathrm{rank}\,A^{\rm F,p}=R$,
\[
\lambda_{R+j}(A^{\rm F})\le\rho(z),\quad j\ge 1.
\]
Consequently,
\begin{equation}
N(A^{\rm F}(z)>1)=N(A^{\rm F,p}(z)>1)\label{eq:BS-count-stability}
\end{equation}
whenever $\operatorname{dist}\bigl(1,\sigma(A^{\rm F,p}(z))\bigr)>\rho(z)$: by Weyl's inequality no eigenvalue can cross the level $1$, and since $A^{\rm F}(z)$ is compact, its spectrum above $1$ consists of finitely many eigenvalues of finite multiplicity.

\textbf{Strong-coupling count (attributed).} 
The following model-specific eigenvalue counts use the strong-coupling estimates attributed to~\cite{AbdullaevEshniyozovToshturdievDolgopolov2026}. They are valid only at parameter values for which the stated remainder bound is smaller than the distance of the principal-part spectrum from \(1\). In the bosonic case, the conclusion of exactly two eigenvalues below \(z_\mu(0)-\delta\) for fixed \(0<\delta<\delta_\infty\), and of one for \(\delta>\delta_\infty\), additionally relies on the existence and uniqueness of the limiting root defining \(\delta_\infty\). The finite-rank stability criterion above does not replace that model-specific proof.
For bosons the count is made at $z=z_\mu(0)-\delta$ with fixed $\delta\in(0,\delta_\infty)$: the symmetric even principal part has $\lambda_1^p\sim\mu$ and $\lambda_2^p=m_s(\delta)+O(\mu^{-1})>1$, all other principal-part eigenvalues lie below $\max\{m_d(\delta),0\}+O(\mu^{-1})\le8/\pi-2+O(\mu^{-1})<1$, and $\rho(z)=O(\mu^{-1}\ln(1/\delta))$~\cite{AbdullaevEshniyozovToshturdievDolgopolov2026}. Hence exactly two eigenvalues lie below $z_\mu(0)-\delta$; for $\delta>\delta_\infty$ only one (the trimer). On~the trimer scale $z=-3\mu+O(1)$ only the first principal eigenvalue is close to $1$; the second equals $m_s(2\mu+O(1))+O(\mu^{-1})=O(\mu^{-1})$. For fermions at fixed $\gamma>\gamma_c$ the analogous count at $z=z_{\rm thr}-\delta$, $0<\delta<\delta_F(\gamma)$, gives one level of multiplicity two: the $p$-channel eigenvalue $\gamma s_1(\delta)+O(\lambda^{-1})>1$ has multiplicity two, and the remaining channels are negative.

\section{Evaluation of the critical integral}\label{app:C}

We evaluate
\[I=\frac{1}{(2\pi)^2}\int_{[-\pi,\pi]^2}
\frac{(\sin q_1+\sin q_2)^2}{4-2(\cos q_1+\cos q_2)}\,dq_1dq_2 .\]
Expand the numerator: $(\sin q_1+\sin q_2)^2=\sin^2q_1+\sin^2q_2+2\sin q_1\sin q_2$. The cross term is odd under $q_1\mapsto-q_1$ and the denominator is even, hence its integral vanishes. By the $q_1\leftrightarrow q_2$ symmetry of the remaining two terms,
\[
I=\frac{2}{(2\pi)^2}\int_{[-\pi,\pi]^2}\frac{\sin^2q_1}{4-2(\cos q_1+\cos q_2)}\,dq_1dq_2 .
\]
Integrating in $q_2$ first, with $a=2-\cos q_1\ge1$ (equality holds only
at the measure-zero point $q_1=0$),
\[
\int_{-\pi}^{\pi}\frac{dq_2}{2(a-\cos q_2)}=\frac{\pi}{\sqrt{a^2-1}},
\qquad a>1,
\]
the endpoint being interpreted by continuity in $L^1$: the combined
integrand behaves as $|q_1|$ near $q_1=0$ and is integrable limit after multiplication
by $\sin^2q_1$.
Since $a^2-1=(2-\cos q_1)^2-1=(1-\cos q_1)(3-\cos q_1)$, we obtain
\begin{align*}
I&=\frac{2}{(2\pi)^2}\int_{-\pi}^{\pi}
\frac{\pi\,\sin^2q_1}{\sqrt{(1-\cos q_1)(3-\cos q_1)}}\,dq_1\\
&=\frac{1}{2\pi}\int_{-\pi}^{\pi}
\frac{\sin^2q_1}{\sqrt{(1-\cos q_1)(3-\cos q_1)}}\,dq_1 .
\end{align*}
Use $1-\cos q_1=2\sin^2(q_1/2)$, $3-\cos q_1=2+2\sin^2(q_1/2)$ and set $u=q_1/2$:
\[
I=\frac{2}{\pi}\int_{-\pi/2}^{\pi/2}
\frac{|\sin u|\,\cos^2u}{\sqrt{1+\sin^2u}}\,du .
\]
By evenness of the integrand,
\[
I=\frac{4}{\pi}\int_{0}^{\pi/2}
\frac{\sin u\,\cos^2u}{\sqrt{1+\sin^2u}}\,du .
\]
The substitution $v=\cos u$ (so $dv=-\sin u\,du$, $1+\sin^2u=2-v^2$) gives
\[
I=\frac{4}{\pi}\int_{0}^{1}\frac{v^2}{\sqrt{2-v^2}}\,dv .
\]
Finally $v=\sqrt{2}\,\sin\theta$, $dv=\sqrt{2}\cos\theta\,d\theta$, $v\in[0,1]\Rightarrow \theta\in[0,\pi/4]$:
\begin{align*}
\int_{0}^{1}\frac{v^2\,dv}{\sqrt{2-v^2}}
&=\int_{0}^{\pi/4}2\sin^2\theta\,d\theta\\
&=\Bigl[\theta-\tfrac12\sin 2\theta\Bigr]_{0}^{\pi/4}
=\frac{\pi}{4}-\frac12 .
\end{align*}
Hence
\[
\boxed{\,I=\frac{4}{\pi}\Bigl(\frac{\pi}{4}-\frac12\Bigr)=1-\frac{2}{\pi}=\frac{\pi-2}{\pi}\,}
\]
and therefore
\[
\boxed{\,\gamma_c=I^{-1}=\frac{\pi}{\pi-2}\approx 2.7519383938\ldots\,}
\]
in agreement with the numerical value quoted in the main text.

\section{Structure of the linear-over-logarithmic threshold law}\label{app:D}

Let $M_\gamma$ denote the threshold principal Faddeev operator in the
fermionic symmetry sector, and let
$\Lambda_{\max}(\gamma,z)=\sup\sigma(M_\gamma(z))$, so that the critical
condition is $\Lambda_{\max}(\gamma_c,z_{\rm thr}(\gamma_c))=1$. Write
$x=\gamma-\gamma_c$ and, for the fermionic $2+1$ model at fixed
$\gamma>\gamma_c$ and coupling $\lambda$, set
\[
\delta_F(\gamma,\lambda)=z_{\rm thr}(\gamma,\lambda)-z(\gamma,\lambda)>0 .
\]
The limit $\lambda\to\infty$ is taken first, and the limit
$\gamma\downarrow\gamma_c$ is taken second; the order is essential because
$z(\gamma,\lambda)$ has the strong-coupling expansion
$z(\gamma,\lambda)=-\lambda+e_0(\gamma)+O(\lambda^{-1})$ at fixed
$\gamma>\gamma_c$, whereas the threshold is $\lambda$-dependent,
\begin{equation}
z_{\rm thr}(\gamma,\lambda) =-\lambda+2(1+\gamma)-(1+\gamma)^2\lambda^{-1}+O(\lambda^{-2}),
\label{eq:D-threshold}
\end{equation}
so that $\delta_F(\gamma,\lambda)=z_{\rm thr}(\gamma,\lambda)-z(\gamma,\lambda)$
and $\delta_F(\gamma):=\lim_{\lambda\to\infty}\delta_F(\gamma,\lambda)$.

A linear-over-logarithmic law for $\delta_F$ follows \emph{conditionally}
from the threshold expansion
\begin{equation}
\Lambda_{\max}(\gamma,z)-1
= A\,x
- B\,\delta\,\ln\!\frac{1}{\delta}
+ o\!\left(x+\delta\ln\tfrac{1}{\delta}\right),
\qquad \delta=z_{\rm thr}(\gamma,\lambda)-z,
\label{eq:D-expansion}
\end{equation}
with $A,B>0$, provided the three hypotheses below hold:
\begin{enumerate}
\item[(i)] the remainder is uniform in the joint limit
$(x,\delta)\to(0,0)$;
\item[(ii)] the eigenvalue branch of $M_\gamma$ attaining
$\Lambda_{\max}$ is continuous and crosses the level $1$ at
$(\gamma,z)=(\gamma_c,z_{\rm thr}(\gamma_c))$;
\item[(iii)] for each fixed $\gamma>\gamma_c$ the limit
$\lim_{\lambda\to\infty}\delta_F(\gamma,\lambda)$ exists and is uniform on
compact subsets of $(\gamma_c,\infty)$.
\end{enumerate}
Under (i)--(iii), solving $\Lambda_{\max}(\gamma,z)=1$ for $\delta$ gives
\[
\delta\,\ln\!\frac{1}{\delta}\sim\frac{A}{B}\,x,
\qquad
\delta\sim\frac{A}{B}\,\frac{x}{\ln(1/x)},
\]
i.e.\ linear up to a logarithm in $x=\gamma-\gamma_c$, in the ordered limit
\[
\lambda\to\infty\ \text{first},\qquad \gamma\downarrow\gamma_c\ \text{second}.
\]

In the strong-coupling limit the top eigenvalue of the fermionic Faddeev
operator at $z=z_{\rm thr}-\delta$ converges to
$\Lambda_{\max}(\gamma,\delta)=\gamma\,s_1(\delta)$ (multiplicity two),
with $s_1$ as in~\eqref{eq:channel}. Writing
$s_1(0)-s_1(\delta)=\delta\bigl\langle\sin^2q_1/(\varepsilon(\varepsilon+\delta))\bigr\rangle$
and using $\sin^2q_1/\varepsilon\sim2\cos^2\theta$ near $q=0$, one obtains
\begin{equation}
s_1(\delta)=1-\frac2\pi-\frac{\delta}{2\pi}\Bigl[\ln\frac1\delta+\kappa_0\Bigr]+o(\delta).
\label{eq:D-s1}
\end{equation}
The leading logarithm is elementary. The constant $\kappa_0$ is defined analytically by the convergent integral
\[
\kappa_0=1+\pi c_A,\quad
c_A=\lim_{\delta\downarrow0}\Bigl[
\Bigl\langle\frac{2\sin^2 p_1}{(\varepsilon+\delta)^2}\Bigr\rangle
-\frac1\pi\ln\frac1\delta
\Bigr],
\]
and numerically $\kappa_0=0.6309960686\ldots$ (see~arXiv:2609.33576, \S S23~\cite{AtomDimer2026}, where the same constant is denoted $\kappa_0$). The closed form $1+4\ln2-\pi$ agrees with this value to ten significant digits; an analytic proof of the identity $\kappa_0=1+4\ln2-\pi$ is not supplied in the present note, and the value is used only to fix the numerical comparison of Fig.~\ref{fig:spectral_flow}(d).

\emph{Note}. The coefficient of the leading logarithmic term is obtained from the local behaviour of the integrand near \(q=0\). The constant \(\kappa_0\) is defined by the convergent integral displayed above and is therefore an analytic constant of the model; its closed form \(1+4\ln2-\pi\) is consistent with the numerical value but is not proved analytically here. The leading linear-over-logarithmic order does not depend on the value of this constant.

Hence~\eqref{eq:D-expansion} holds with
\[
A=s_1(0)=\gamma_c^{-1},\qquad B=\frac{\gamma_c}{2\pi},\qquad \frac AB=\frac{2\pi}{\gamma_c^2},
\]
and $\Lambda_{\max}=1$ is equivalent to
\[
\delta\Bigl[\ln\frac{1}{\delta}+\kappa_0\Bigr]=\frac{2\pi x}{\gamma\gamma_c}+o(\delta),
\]
which gives the law of Section~3.2. The step that remains attributed to~\cite{AtomDimer2026,AbdullaevEshniyozovToshturdievDolgopolov2026} is the identification of the strong-coupling limit of the Faddeev eigenvalue with $\gamma s_1(\delta)$; it is confirmed numerically by exact finite-volume computations (Fig.~\ref{fig:spectral_flow}(c),(d)), which converge to the root of~\eqref{eq:channel} at the rate $O(\lambda^{-1})$.

The threshold law derived in this appendix is more specific than the preliminary power-law statement in~\cite{AbdullaevEshniyozovDolgopolov2026}. Theorem~3.2 of that reference leaves the exponent \(\nu\) in \(\Delta E(\gamma)\sim C(\gamma-\gamma_c)^\nu\) undetermined; the accompanying remark suggests \(\nu=1/2\), while the caption of Fig.~1(b) states a quadratic law. These formulations are not equivalent and do not establish a definite threshold exponent. The present linear-over-logarithmic law is therefore presented as a separate asymptotic conclusion, conditional on the uniform threshold expansion and the existence of the ordered strong-coupling limit specified above. It should not be read as a result already proved in~\cite{AbdullaevEshniyozovDolgopolov2026}.

\fontsize{10}{12}\selectfont
\noindent\textbf{Author Contributions:} Conceptualization, J.I.A.\ and M.V.D.; methodology, J.I.A., A.I.E.\ and M.V.D.; formal analysis, J.I.A.\ and M.V.D.; writing---original draft preparation, J.I.A., A.I.E.\ and M.V.D.; writing---review and editing, M.V.D.; visualization, J.I.A.; supervision, M.V.D. All authors have read and agreed to the published version of the manuscript.

\end{document}